\documentstyle[12pt, amsmath, amsfonts, epsfig, amssymb, rotating]{article}
\begin{document}
\def\be{\begin{equation}}
\def\eps{\epsilon}
\def\cala{{\mathcal A}}
\def\calm{{\mathcal M}}
\def\calb{{\mathcal B}}
\def\calc{{\mathcal C}}
\def\calh{{\mathcal H}}
\def\rm{\mathbb{R}}
\def\nm{\mathbb{N}}
\def\cm{\mathbb{C}}
\def\zm{\mathbb{Z}}
\def\hm{\mathbb{H}}
\def\pd{\pi/2}
\def\ua{\underline{a}}
\def\ub{\underline{b}}
\def\tr{\textrm{tr}}
\def\im{\textrm{Im}}
\def\Ul{\overleftarrow{U}}
\def\Ur{\overrightarrow{U}}
\def\Dl{\overleftarrow{D}}
\def\Dr{\overrightarrow{D}}

%%%%%%%%%%%%%%%%%%%%%%%%%% Format de la page %%%%%%%%%%%%%%%%%%%%%%%%%%%%%%%%%%

\textwidth= 16cm
\oddsidemargin= 0.5cm
\evensidemargin=-0.5cm
\topmargin=-1cm
\textheight= 21cm

\title{\bf Diffractive orbits in isospectral billiards}
\author{O. Giraud\\
H. H. Wills Physics Laboratory\\
Tyndall Avenue\\
Bristol BS8 1TL, UK}

\maketitle

\begin{abstract}

Isospectral domains are non-isometric regions of space for which the spectra
of the Laplace-Beltrami operator coincide. In the two-dimensional Euclidean space, 
instances of such domains have been given. It has been proved for these examples that the
length spectrum, that is the set of the lengths of all periodic trajectories, coincides as well.
However there is no one-to-one correspondence between the diffractive trajectories. 
It will be shown here how the diffractive contributions to the Green functions match nevertheless
in a ``one-to-three'' correspondence.

\end{abstract}

\pagebreak

\section{Introduction}

The quantum-mechanical problem of finding isospectral domains, that
is two non-isometric regions for which the sets $\{E_n, n\in\nm\}$ of solutions of the 
stationary Schr\"odinger equation
\begin{equation}
\label{helmholtz}
(\Delta+E)\Psi=0,
\end{equation}
with $\Psi|_{\textrm{boundary}}=0$,
are identical, has been formulated in a synthetic way by Mark Ka{\v c} in 1966 in his famous paper
``Can one hear the shape of a drum'' \cite{Kac66}. Negative answers to this problem have been given
for domains on Riemannian manifolds \cite{Be89, Br88}. The answer for two-dimensional Euclidean domains
was finally given
by Gordon {\it et al.} \cite{GorWebWol92}, who provides explicitly a pair of simply 
connected non-isometric Euclidean isospectral domains. The two billiards considered in 
\cite{GorWebWol92} are represented in Figure \ref{chapman}a.
Using a paper-folding method, Chapman \cite{Cha95} showed that in this case  
isospectrality arises from the existence of a map between the two domains. This method allowed him to
produce more examples of isospectral billiards (by billiard we mean a two-dimensional Euclidean 
connected compact domain): he showed that any triangle or even rectangle could replace the base right angled
isosceles triangle used to build the billiards in \ref{chapman}a; one just has to glue 
together 7 copies of the chosen base shape obtained by symmetry with respect to its edges, in the
same order as to build the billiards of \ref{chapman}a. We will work here with the pair of billiards 
$\calb_{1}$ and $\calb_{2}$ of Figure \ref{chapman}b, where each billiard is made of seven rectangles.
Later on, Buser {\it et al} \cite{BusConDoy94} produced more examples of planar isospectral domains made 
of more building blocks; all of them are based on the same principle of gluing together copies of a 
base triangle.

\begin{figure}[ht]
\begin{center}
\epsfig{file=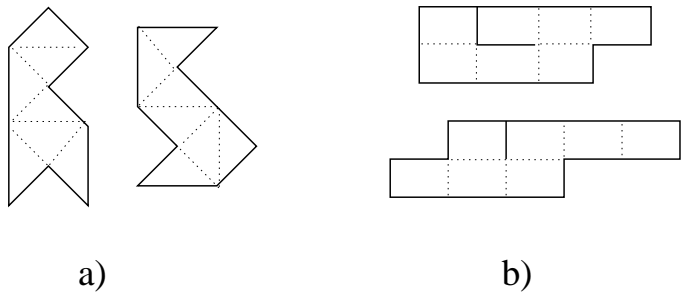,width=9cm}
\end{center}
\caption{a) Two isospectral billiards with a triangular base shape; b) The same with a rectangular base shape.}
\label{chapman}
\end{figure}

A natural question arises when one considers isospectral billiards: is there any relation between their
periodic orbits, and is there any relation between their diffractive orbits?
It is well-known that the quantum density of states
\begin{equation}
d(E)=\sum_n\delta(E-E_n)
\end{equation}
can be expressed by means of the advanced Green function as
\begin{equation}
\label{densitegreen}
d(E)=-\frac{1}{\pi}\im\int d\ua G(\ua, \ua),
\end{equation}
where the integral is performed over the domain.
In the case of two isospectral billiards, one might naturally expect the integrals of the Green functions
of the two systems to be equal. Since the Green function can be expanded as a sum over all Feynman 
paths, the correspondence between the spectra should be associated to a correspondence between 
the periodic orbits and between the diffractive orbits.
Moreover, there exists an exact expression for the Green function in a two-dimensional polygonal billiard: following
Sommerfeld \cite{Som96, Som54}, who provides the exact Green function for a half infinite straight mirror in 2
dimensions, Stovicek \cite{Sto89, Sto91b} expressed the Green function for a collection of magnetic flux lines 
on a plane (the multi-flux Aharonov-Bohm effect) as a sum over all possible scattering paths. This
method has been generalized in \cite{HanTha03} to provide the exact Green function for the scalar 
wave equation in a plane with any set of perfectly reflecting straight mirrors joined by diffractive corners.
The Green function is given as a  scattering series involving all classical trajectories
and all scattering contributions; a semi-classical series expansion shows that one expects the Green
functions of the two system to correspond order by order.

The correspondence between the periodic orbits of two isospectral billiards
(which contribute to the lowest order of the scattering series)
has been discussed in \cite{Gor86} for isospectral domains on Riemannian 
manifolds. 
The Laplace spectrum versus the length spectrum is also discussed by Gordon in \cite{GorWebWol92}
for Euclidean isospectral billiards. A proof of the one-to-one correspondence between 
the length spectra (referred to as ''iso-length spectrality'') in the case of the two celebrated 
isospectral billiards considered by Gordon is given in \cite{Tha03}, based on simple mathematical 
tools. More generally, iso-length spectrality has been proved in \cite{OkaShu01} for all pairs of isospectral billiards
having a ''transplantation'' property (which corresponds, roughly speaking, to the existence
of a map between the two domains, such as the one that will be given in Section \ref{formalism}).
However, what  \cite{Tha03}
underlines is that there is no  one-to-one correspondence between 
the diffractive trajectories of the two billiards. More precisely, one can find straight lines between
diffractive corners in one of the billiards that do not have any counterpart in the other. Still, the
equality of the spectra, and hence the equality of the trace of the Green function, imposes
some correspondence.
In Section \ref{deux} we will study the map that exists between the two billiards and show how this
bijection allows to prove a one-to-one correspondence of the periodic orbits. In Section \ref{trois}
we establish a relation between the Green function of the two studied domains. Finally in Section \ref{quatre}
we answer the following question: what is the correspondence between the diffractive orbits.
The conclusion briefly shows how the method presented here can be naturally extended
to other pairs of isospectral billiards.

\section{Isospectrality and periodic orbits}
\label{deux}

\subsection{The translation surfaces}
\label{formalismematriciel}

Instead of studying directly the billiards $\calb_1$ and $\calb_2$ of Figure \ref{chapman}b, we 
will consider the equivalent problem of studying the translation surfaces \cite{GutJud00} associated to 
these billiards. $\calb_1$ and $\calb_2$ 
are polygons with angles ($\pd$, $\pd$, $\pd$, $\pd$, $\pd$, $\pd$, $\pd$, $3\pd$, $3\pd$, $2\pi$). 
A construction due to Zemlyakov and Katok \cite{ZemKat76} shows that the translation surface 
associated to a generic rational polygonal billiard  is obtained by unfolding the polygon 
with respect to each of its sides, which means gluing to the initial polygon its images by reflexion 
with respect to each of its sides and repeating the operation. If $\alpha_i=\pi m_i/n_i$ are the angles
of the polygon and $N$ the least common multiple of the $n_i$, then $2N$ copies of the initial billiard
are needed. In the case of the billiards $\calb_{1}$ and $\calb_{2}$, since all the angles are multiples
of $\pi/2$,  only 4 copies are needed,
and the translation surfaces $\calm_{1}$ and $\calm_{2}$ obtained by this construction are represented 
in Figure \ref{structures}. In these surfaces, all opposite sides are identified.
Note that each structure has 4 singularities: 2 angles of measure $6\pi$ (dot and circle), and 2 angles 
of measure $4\pi$ (cross and star). Moreover, each singularity with angle $2(k+1)\pi$ brings the
contribution $k$ to the quantity $2g-2$, where $g$ is the genus of the surface \cite{BerRic81}. Therefore
the surfaces $\calm_{1}$ and $\calm_{2}$ both are of genus 4. Any path drawn on $\calm_\nu$ corresponds
to an unfolded path on $\calb_\nu$.
\begin{figure}[ht]
\begin{center}
\epsfig{file=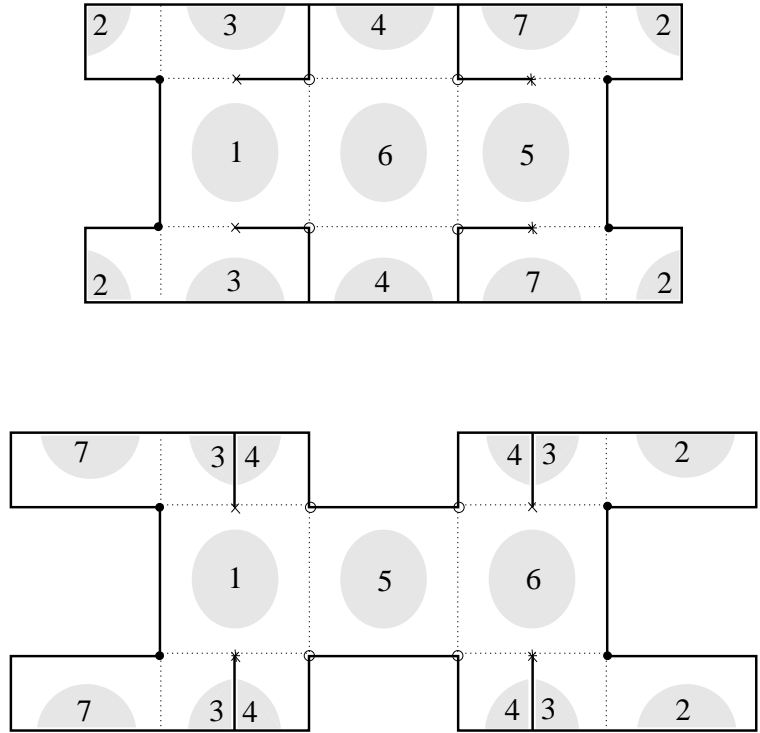,width=10cm}
\end{center}
\caption{4 copies of the two isospectral billiards $\calb_1$ and $\calb_2$ are glued together
to make two isospectral translation surfaces $\calm_1$ and $\calm_2$ made from 7 tiles.}
\label{structures}
\end{figure}

Each surface is made of 7 tiles, which give to each translation surface the structure of a 7-fold 
torus cover. If the tiles are numbered as in Figure \ref{structures}, the way these tiles
are glued together to form a surface of genus 4 can be expressed, following \cite{Tha03}, 
through three $7\times 7$ matrices: for
each structure $\calm_{\nu},\ \nu=1,2$, we introduce the matrices $R^{(\nu)}$, $\Ul^{(\nu)}$ and 
 $\Ur^{(\nu)}$ such that $R^{(\nu)}_{i,j}=1$ if the right edge of tile $i$ is glued to 
the left edge of tile $j$ and 0 otherwise,
$\Ul^{(\nu)}_{i,j}=1$ if the left half of the upper edge of tile $i$ is glued to 
the left half of the lower edge of tile $j$ and 0 otherwise,
and of course $\Ur^{(\nu)}_{i,j}=1$ if the right half of the upper edge of tile $i$ is glued to 
the right half of the lower edge of tile $j$ and 0 otherwise. 
Just looking at Figure \ref{structures}, we can obtain these matrices easily. They are
given in Appendix \ref{matricesRUU}. For instance, in $\calm_1$, the right neighbour 
of tile 5 is tile 1, the right neighbour of tile 3 is tile 7, etc...
We can also define the matrices $L^{(\nu)}$, $\Dl^{(\nu)}$ and
$\Dr^{(\nu)}$ which indicate which tile is glued to the left, the bottom left or the bottom right of
a given tile. These three matrices are nothing but the transposes of respectively $R^{(\nu)}$, $\Ul^{(\nu)}$ and 
 $\Ur^{(\nu)}$.

\subsection{Mapping between translation surfaces}
\label{formalism}

The isospectrality of the two billiards arises from the existence of a mapping between $\calm_{1}$ and 
$\calm_{2}$, provided by Gordon \cite{GorWebWol92} and made explicit by Chapman \cite{Cha95}. The
following section proves the isospectrality between $\calm_1$ and $\calm_2$.

By convention, we will label any point $\ua$ in $\calm_1$ or $\calm_2$ (which are made of 7 tiles
ore tori) by its position $a$ in the
tile and the number $i$ ($1\leq i \leq 7$) of the tile it is in; we will write alternatively $\ua$ or
$(a,i)$. Let us define the ''transplantation matrix'' $T$ as
\begin{equation}
\label{matriceT}
T=\left(
\begin{array}{ccccccc} 1&0&0&1&0&0&1\cr 0&1&0&0&1&0&1\cr 0&0&1&0&0&1&1\cr
1&0&0&0&1&1&0\cr 0&1&0&1&0&1&0\cr 0&0&1&1&1&0&0\cr 1&1&1&0&0&0&0\cr\end{array}\right).
\end{equation}
The isospectrality arises from the fact that for any given eigenstate $\phi_n$ of $\calm_1$, we can construct 
an eigenstate $\psi_n$ in $\calm_2$ defined by 
\begin{equation}
\label{psiphi}
\psi_n(a, i)=\frac{1}{\cala_n}\sum_j T_{i, j}\phi_n(a,j),
\end{equation}
where $\cala_n$ is a normalization factor that we will discuss in Section \ref{trois}.
For instance, for $a$ in tile 1 of $\calm_2$, we have
\begin{equation}
\psi_n(a,1)=\frac{1}{\cala_n}\left(\phi_n(a,1)+\phi_n(a,4)+\phi_n(a,7)\right).
\end{equation}
We call the tiles 1,4,7 in $\calm_1$ the ''pre-images'' of tile 1 in $\calm_2$, and we say that 1 
is ''made of'' tiles 1,4 and 7.
The fact that the functions $\psi_n$ are the eigenstates of
$\calm_2$ comes from the following relations:
\begin{equation}
\label{commutation}
R^{(2)}T=TR^{(1)},\ \  \Ul^{(2)}T=T\Ul^{(1)},\ \  \Ur^{(2)}T=T\Ur^{(1)},
\end{equation}
which can be verified directly on the matrices given in Appendix \ref{matricesRUU}.
Each of these commutation relation has a natural interpretation. For instance the 
commutation relation for $R$ means that for any pair $(i,j)$ of tiles, $i$ on $\calm_2$ and $j$
on $\calm_1$, we have
\begin{equation}
\label{krt}
\sum_{k}R^{(2)}_{i,k}T_{k,j}=\sum_{k'}T_{i,k'}R^{(1)}_{k',j},
\end{equation}
which means that
if $k$ is the tile on the right of tile $i$ in $\calm_2$ (i.e. $R^{(2)}_{i,k}=1$) and if we call
$i_1, i_2, i_3$ the pre-images of $i$ (i.e. $T_{i,i_1}=T_{i,i_2}=T_{i,i_3}=1$), then
by (\ref{krt})
\be
\label{trr}
T_{k,j}=R^{(1)}_{i_1,j}+R^{(1)}_{i_2,j}+R^{(1)}_{i_3,j}.
\end{equation}
Among the 7 possible values of $j$, the right-hand side of Equation (\ref{trr}) will be 1
if and only if $j$ is on the right of $i_1$, $i_2$ or $i_3$ in $\calm_1$, and 0 otherwise.
This implies that if $k$ is on the right of $i$,
the three pre-images $k_1, k_2, k_3$ of $k$ (verifying $T_{k,k_i}=1$)
are the three tiles on the right of
the three pre-images of $i$. Concretely, since tile 1 in $\calm_2$ is made of tiles 1,4,7, then
its right neighbour tile 5 is made of the right neighbours 2,4,6 of tiles 1,4,7. Since the
commutation relation (\ref{commutation}) is valid for all matrices $A\in\{R,L,\Ur,\Ul,\Dr,\Dl\}$
all properties of continuity between tiles in $\calm_1$ are preserved by transplantation.
Therefore all the functions $\psi_n$ constructed by (\ref{psiphi}) are continuous on 
$\calm_2$. Obviously these functions verify the Helmholtz equation (\ref{helmholtz})
as a linear combination of solution, and with the same eigenvalues. Finally, one has to
note that since $T$ is an invertible matrix,
only $\phi_n=0$ would give $\psi_n=0$. This proves the isospectrality between $\calm_1$
and $\calm_2$.

\subsection{Trajectories on the translation surfaces}
\label{principeessentiel}

The matrix formalism introduced in subsection \ref{formalismematriciel} allows us to construct
a ''movement matrix'' $M$ from any path drawn on the translation surfaces $\calm_1$ or $\calm_2$, 
adapting the method explained in \cite{Tha03}. 
Each tile has 6 neighbours, and any path is drawn on a sequence $(i_1,i_2,...,i_n)$ of tiles
such that $i_{k+1}$ is a neighbour of $i_k$, provided it does not hit any vertex (we call ''vertex''
a point on the surface where 4 tiles join, or a point in the middle of a horizontal edge; 
there are 4 scattering vertices, the others are non-scattering ones).
Let us call $A_k\in\{R,L,\Ur,\Ul,\Dr,\Dl\}$ the matrix corresponding
to the movement from $i_k$ to $i_{k+1}$, i.e. the matrix verifying $(A_k)_{i_k, i_{k+1}}=1$.
If we define the movement matrix $M$ as $\prod_{k=1}^{n} A_k$, it will verify $M_{i_1, i_n}=1$. 
As a product of permutation matrices,
$M$ is a permutation matrix, and it maps tile $i_1$ onto tile $i_n$.
For instance the path drawn in Figure \ref{exemplepath}  
corresponds to a sequence of tiles $(1,6,4,4,6,5)$ and to a sequence of matrices $M=R\Ul L \Ur R$.
\begin{figure}[ht]
\begin{center}
\epsfig{file=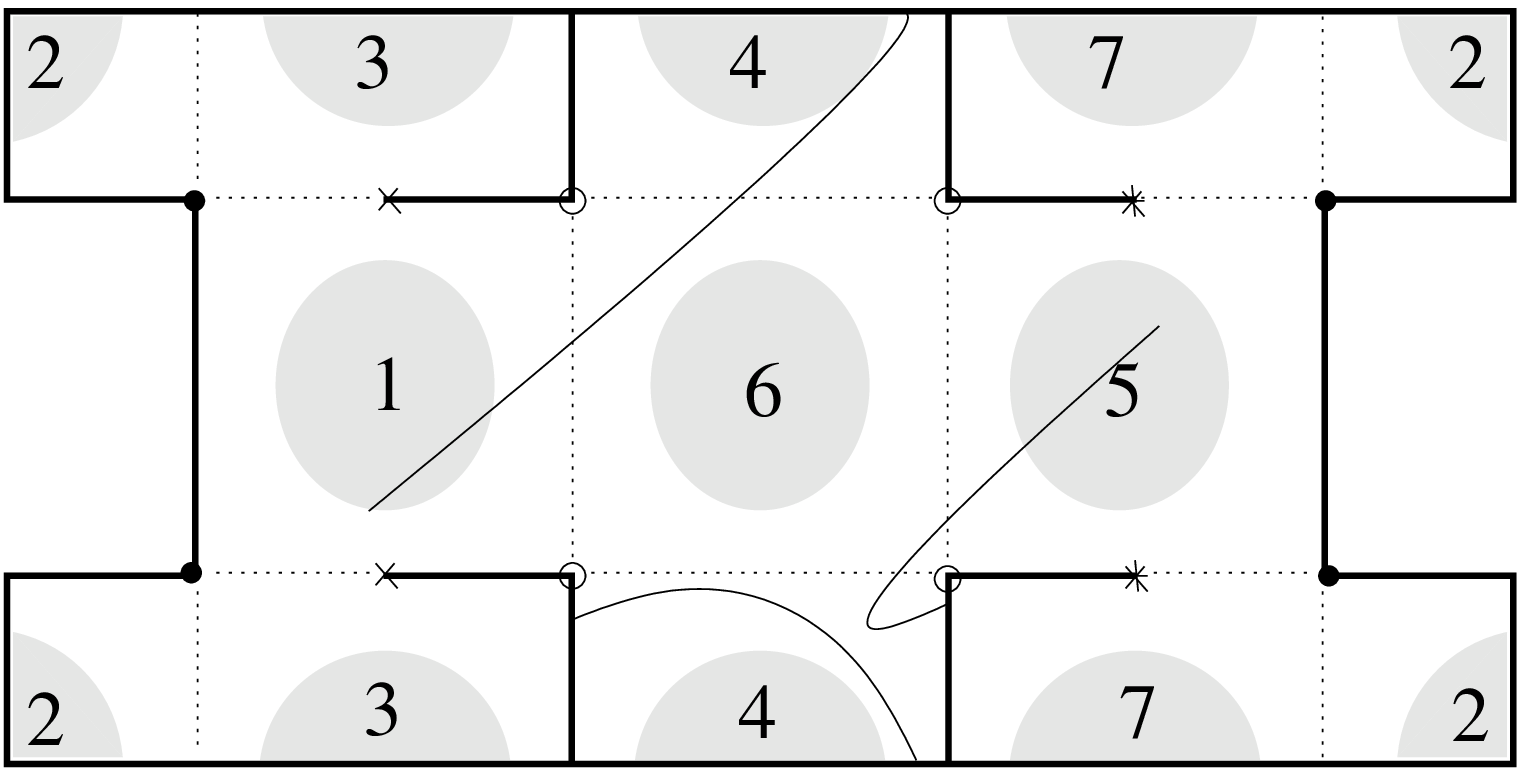,width=7cm}
\end{center}
\caption{A path drawn on $\calm_1$}
\label{exemplepath}
\end{figure}
Reciprocally, for any product $M=A_1 A_2...A_n$ of matrices belonging
to $\{R,L,\Ur,\Ul,\Dr,\Dl\}$, if $M_{i, j}=1$ then there is a sequence $(i_1=i,i_2,...,i_n=j)$ 
of tiles such that $i_{k+1}$ is a neighbour of $i_k$ and $(A_k)_{i_k, i_{k+1}}=1$. Note
that infinitely many sequences  $(i_1,i_2,...,i_n)$ give rise to the same movement matrix,
because there is an infinite number of sequences and only a finite number (7!=5040) of 
permutations.
\\
\\
Let us now consider a sequence of matrices $(A^{(1)}_1, A^{(1)}_2,..., A^{(1)}_n)$, and the movement matrix 
$M^{(1)}\equiv A^{(1)}_1 A^{(1)}_2... A^{(1)}_n$; then for any $i$
it defines a unique sequence of tiles $(i=i_1, i_2,...,i_n)$ such that $(A^{(1)}_k)_{i_k, i_{k+1}}=1$.
This sequence has the property that $M^{(1)}_{i_1, i_n}=1$. According to the relation 
(\ref{commutation}), each $A^{(1)}_k$ verifies $A^{(1)}_k=T^{-1} A^{(2)}_k T$, which implies that 
the matrix $M^{(2)}\equiv A^{(2)}_1 A^{(2)}_2... A^{(2)}_n$ verifies
\begin{equation}
\label{commutationM}
T M^{(1)}= M^{(2)} T.
\end{equation}
The interpretation of this commutation relation is the same as the interpretation of (\ref{commutation}) for the 
individual $A_k$ (see Equations (\ref{krt}) and  (\ref{trr}) with $R$ replaced by $M$):
 if one can go from $i_1$ to $i_n$ by a sequence of tiles glued in a specific way, then
 the three pre-images of $i_n$ are the tiles obtained by starting from the three pre-images of $i_1$
and following a sequence of tiles glued in exactly the same way.
To any path drawn on $\calm_2$ not hitting corners it is possible to associate the sequence of tiles
on which it is drawn, starting from a tile $i$ and finishing on a tile $j$; it is therefore  
possible to draw an identical path starting from any of the three pre-images of $i$; this path will 
necessarily arrive to one of the three pre-images of $j$.

\subsection{Equality of the length spectrum}

The two surfaces 
$\calm_1$ and $\calm_2$ are pseudo-integrable, therefore their periodic orbits occur in families
of parallel orbits of same length \cite{BerRic81}.
It has been shown in \cite{Tha03} that there is an exact one-to-one correspondence between 
the pencils of periodic orbits of $\calm_1$ and those of $\calm_2$. The main argument is the
principle explained in section \ref{principeessentiel} that for any path drawn on one of
the surfaces there is a corresponding sequence of tiles and a
corresponding sequence of matrices. The product of these matrices, the ''movement matrix'' $M$,
has the property that $M_{i,j}=1$ if and only if the sequence of tiles goes from $i$ to
$j$. For any periodic orbit on $\calm_1$ the last tile has to be equal to the first one, and
a closed path going from tile $i$ to itself has a movement matrix verifying
$M^{(1)}_{i,i}=1$. The quantity tr$(M^{(1)})$ is therefore the number of tiles from which
one can start and come back to oneself after a sequence of tiles giving the movement matrix 
$M^{(1)}$.
Since the commutation relation (\ref{commutationM}) implies that $\tr(M^{(1)})=\tr(M^{(2)})$,
there is the same number of tiles having this property in $\calm_2$. So for any periodic pencil
drawn on this sequence of tiles there will be  tr$M^{(1)}$ identical copies of it 
(in particular with same length and same width) on $\calm_1$
and the same number on $\calm_2$, hence the bijection between the periodic orbits.
As an illustration, Figure \ref{2op} shows the two pencils of periodic orbits in a given 
direction on $\calm_1$ and $\calm_2$: the grey orbit has same length and same width on both
surfaces, and so does the white orbit.
\begin{figure}[ht]
\begin{center}
\epsfig{file=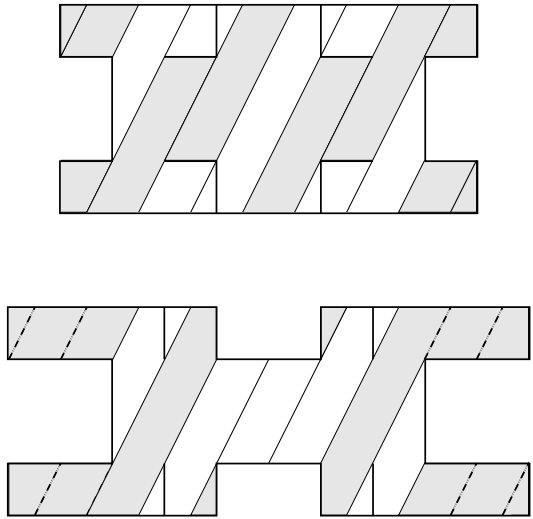,width=7cm}
\end{center}
\caption{Two identical strips of periodic orbits on the surfaces $\calm_1$ and $\calm_2$}
\label{2op}
\end{figure}

\section{The Green function}
\label{trois}

Relation (\ref{densitegreen}) implies that the imaginary part of the trace of the (retarded)
Green function is identical for two isospectral billiards. Here we will be more precise and
express the Green function of $\calm_2$ in terms of the Green function of $\calm_1$.

\subsection{Tile modes and normalization}

Each surface $\calm_1$ and $\calm_2$ is made out of 7 tiles glued together. We will call
tile modes the solutions of the Helmholtz equation (\ref{helmholtz}) on a tile with periodic boundary
conditions, that is a torus of size $u\times v$. The eigenvalues corresponding
to these tile modes will be denoted $E_t$, and the corresponding eigenfunctions $\chi_t$. 

There is a one-to-one correspondence between these tile modes $E_t$ and a subset of
the spectrum $E_n$ common to the surfaces $\calm_1$ and $\calm_2$.
First, any eigenfunction $\chi_t$ corresponds to a solution $\psi_t(a,i)$ (or $\phi_t(a,i)$)
of the Helmholtz equation on $\calm_1$ (or $\calm_2$) by simply taking the function equal
to (or proportional to) $\chi_t(a)$ on each tile: the periodic boundary conditions for 
the tile eigenfunctions $\chi_t$ will make $\psi_t$ and $\phi_t$ continuous. In order to normalize
correctly to 1 these eigenfunctions, one has to set 
\begin{equation}
\label{tilepsiphi}
\psi_t(a,i)=\phi_t(a,i)=\frac{1}{\sqrt{7}}\chi_t(a).
\end{equation}
Reciprocally, for any eigenstate $\phi_n$ of the surface $\calm_1$, the function defined on
a tile by
\begin{equation}
\chi_n(a)=\sum_{i=1}^{7}\phi_n(a, i)
\end{equation}
is either the function 0 or an eigenstate $\chi_t$ of the tile. We will use the subscript 
$s$ when the state $\phi_n$ verifies the condition
\begin{equation}
\label{sommezero}
\sum_{i=1}^{7}\phi_n(a, i)=0
\end{equation}
and the subscript $t$ when it does not. The set of eigenstates of $\calm_1$ 
(and, in the same way, the set of eigenstates of $\calm_2$)
can then be partitioned into two sets: 
the eigenstates $\phi_t$ which are also the eigenstates of the tile
and do not  have the property (\ref{sommezero}), and 
the eigenstates $\phi_s$ which have the property (\ref{sommezero}) (pure ''surface'' states).

The normalization constant $\cala_n$ in (\ref{psiphi}) will take a different value whether
the function $\phi_n$ belongs to the set of $\{\phi_t\}$ or  $\{\phi_s\}$, as we will see now.
Equation (\ref{psiphi}) expresses each eigenfunction $\psi_n$ of $\calm_2$ as a sum of
an eigenfunction $\phi_n$ of $\calm_1$ with the same eigenvalue, taken at three
different points of $\calm_1$ (or rather at a similar point on three different tiles).
The normalization condition on $\psi$ can be written
\begin{eqnarray}
1=\int_{\calm_2}\left|\psi_n(x)\right|^2 dx 
&=&\sum_{k=1}^{7}\int_{\textrm{tile}}\left|\psi_n(a,k)\right|^2 da\nonumber\\
&=&\frac{1}{\cala_n^2}\sum_{k=1}^{7}\sum_{i,j=1}^{7}T_{k,i}T_{k,j}\int_{\textrm{tile}}
\overline{\phi_n(a,i)}\phi_n(a,j)da\nonumber\\
&=&\frac{1}{\cala_n^2}\sum_{i,j=1}^{7}T^2_{i,j}\int_{\textrm{tile}}\overline{\phi_n(a,i)}\phi_n(a,j)da
\end{eqnarray}
where we have used the fact that $T$ is symmetric and summed over $k$. Since
the matrix $T^2$ is equal to $(T^2)_{i,j}=1+2\delta_{ij}$ ($\delta_{ij}$ is the
Kronecker symbol), we get
\begin{eqnarray}
\label{norm}
\cala_n^2&=&\sum_{i,j=1}^{7}\int_{\textrm{tile}}\overline{\phi_n(a,i)}\phi_n(a,j)da
+2\sum_{i=1}^{7}\int_{\textrm{tile}}|\phi_n(a,i)|^2da\nonumber\\
&=&\int_{\textrm{tile}}\left|\sum_{i=1}^{7}\phi_n(a,i)\right|^2 da+2\int_{\calm_1}|\phi_n(a,i)|^2 da.
\end{eqnarray}
For the tile modes $\psi_t$, we can replace $\phi$ by $\chi/\sqrt{7}$ from Equation (\ref{tilepsiphi}).
This yields
\begin{equation}
\cala_t^2=\frac{49}{7}\int_{\textrm{tile}}\left|\chi_t(a)\right|^2 da
+\frac{14}{7}\int_{\textrm{tile}}\left|\chi_t(a)\right|^2 da=9
\end{equation}
because the functions $\chi_t$ are normalized to 1 on the tile. For the non-tile modes $\phi_s$, which
verify Equation (\ref{sommezero}), Equation (\ref{norm}) gives $\cala_s^2=2$. Finally we have
\begin{eqnarray}
\label{tilenontile}
\psi_t(a, i)=\frac{1}{3}\sum_j T_{i, j}\phi_n(a,j)&\ \ \ \ \ \ \ \ \textrm{(tile modes)}\nonumber\\
\psi_s(a, i)=\frac{1}{\sqrt{2}}\sum_j T_{i, j}\phi_n(a,j)&\ \ \ \ \ \ \ \ \textrm{(surface modes)}.\nonumber\\
\end{eqnarray}

\subsection{The Green function of the translation surfaces}

In this section, we want to express the Green function on the surface $\calm_2$ in
terms of the Green function of the surface $\calm_1$. We will use the expansion over
eigenstates for the advanced Green function of $\calm_2$:
\begin{equation}
G^{(2)}(\ua,\ub)=\sum_{n}\frac{\overline{\psi_n(\ua)}\psi_n(\ub)}{E-E_n+i\eps}
\end{equation}
where $\psi_n$ and $E_n$ are respectively the eigenfunctions and the eigenvalues of $\calm_2$.
We have to split the sum over $n$ into a sum over the tile modes $\psi_t$ and the non-tile modes 
$\psi_s$, and replace $\psi$ by its expression (\ref{tilenontile}):
\begin{eqnarray}
G^{(2)}(a,i;b,j)
=\frac{1}{9}\sum_t\frac{\sum_{i',j'}T_{i,i'}T_{j,j'}\overline{\phi_t(a,i')}\phi_t(b,j')}{E-E_t+i\eps}
\nonumber\\
+\frac{1}{2}\sum_s\frac{\sum_{i',j'}T_{i,i'}T_{j,j'}\overline{\phi_s(a,i')}\phi_s(b,j')}{E-E_s+i\eps}.
\end{eqnarray}
If we add and subtract $(1/2)\sum_t$, we get
\begin{eqnarray}
G^{(2)}(a,i;b,j)
=\frac{1}{2}\sum_{i',j'}T_{i,i'}T_{j,j'}\sum_n\frac{\overline{\phi_n(a,i')}\phi_n(b,j')}{E-E_n+i\eps}
\nonumber\\
-\frac{7}{18}\sum_{i',j'}T_{i,i'}T_{j,j'}\sum_t\frac{\overline{\phi_t(a,i')}\phi_t(b,j')}{E-E_t+i\eps}.
\end{eqnarray}
The first sum is a sum over 
Green functions of $\calm_1$; in the sum over the remaining $t$ modes, we replace $\phi_t$ 
by its value given by Equation (\ref{tilepsiphi}).
This gives
\begin{equation}
G^{(2)}(a,i;b,j)=\frac{1}{2}\sum_{i',j'}T_{i,i'}T_{j,j'}G^{(1)}(a,i';b,j')
- \frac{1}{18}\sum_{i',j'}T_{i,i'}T_{j,j'}G^{(t)}(a;b)
\end{equation}
where $G^{(t)}(a;b)$ is the Green function on the tile. This Green function 
does not depend on $i'$ or $j'$: therefore
the sum over $i'$ and $j'$ can be performed: $\sum_{i',j'}T_{i,i'}T_{j,j'}=9$ since each row and each column
of $T$ has three 1's. Finally the Green function of the surface $\calm_2$ is
\begin{equation}
\label{relationG}
G^{(2)}(a,i;b,j)=\frac{1}{2}\sum_{i',j'}T_{i,i'}T_{j,j'}G^{(1)}(a,i';b,j')
- \frac{1}{2}G^{(t)}(a;b).
\end{equation}
Obviously, a similar relation can be obtained that expresses $G^{(1)}$ as a sum over functions $G^{(2)}$.

\section{Diffractive orbits}
\label{quatre}

\subsection{Stovicek's formalism}

We know from \cite{HanTha03} that the exact expression for the Green function can be obtained 
for billiards with a polygonal enclosure as a scattering series, where each term is
a sum over scattering paths made of straight lines and scatters on the singular corners. 
Each scattering path contributing to $G(\ua, \ub)$ is made
of a starting leg of length $r_0$ going from the initial point $\ua$ to a scatterer, then an
alternating series of diffractions with angles $\varphi_i\in\rm$ followed by legs of length $r_i$, $1\leq i \leq n$,
the last leg of length $r_n$ going from the last scatterer to the final point $\ub$. The expression given in
\cite{HanTha03} for scattering on straight reflectors can be adapted here and gives
\begin{eqnarray}
\label{greenstovicek}
G(\ua, \ub)&=&\sum_{n=0}^{\infty}\frac{1}{(2\pi)^n}\sum_{\genfrac{}{}{0pt}{}{n\ \textrm{vertex}}{\textrm{paths}}}
\frac{1}{2i}\int_{-\infty}^{\infty}ds_1 ds_2 ... ds_n H_0^{(1)}\left[k R(s_1, s_2, ..., s_n)\right]\nonumber\\
&\times&\prod_{k=1}^{n}\frac{2\pi}{(\gamma_k M_k+\theta_k+i s_k)^2-\pi^2},
\end{eqnarray}
where
\begin{eqnarray}
R^2(s_1, s_2, ..., s_n)=\left(r_0+r_1 e^{s_1}+r_2 e^{s_1+s_2}+\cdots+r_n e^{s_1+s_2+\cdots+s_n}\right)\nonumber\\
\times\left(r_0+r_1 e^{-s_1}+r_2 e^{-s_1-s_2}+\cdots+r_n e^{-s_1-s_2-\cdots-s_n}\right).
\end{eqnarray}
The $k$-th diffraction angle $\varphi_k$ is equal to $M_k \gamma_k+\theta_k$; here $\gamma_k$
is the measure of the angle at the singularity ($4\pi$ or $6\pi$ in our case), 
$M_k$ is the winding number (i.e. the number
of times the path winds around the singularity), and $0\leq \theta_k<\gamma_k$.
A schematic example of such a scattering orbit is provided at Figure \ref{pathinfini}.

\subsection{Diffractive orbits}

Let us now look at the ''saddle-connexions'' \cite{EskMasSch01} which we define as the geodesics joining
two diffracting vertices. If $u$ and $v$ are the width and the height of a tile, 
the vertices (scattering and non-scattering)
are in the directions $(m u/2, n v)$ with $m,n\in\zm$. Each pair $(m,n)$ with $m$ and $n$ co-prime
will give the direction of a saddle-connexion. If we define
\begin{equation}
\label{lp}
l_p=\sqrt{(m u/2)^2+(n v)^2}
\end{equation}
for $p=(m,n)\in {\zm}^2$, $m$ and $n$ co-prime, the lengths of the saddle-connexions will be
of the form $k l_p$, $k\in\nm$. These saddle-connexions will be the legs of the scattering paths
in the expansion (\ref{greenstovicek}). 

But there is no one-to-one correspondence between
the saddle-connexions of $\calm_1$ and those of $\calm_2$. In fact, there are saddle-connexions
in $\calm_2$ that cannot be found anywhere in $\calm_1$. For instance, the dashed saddle-connexion
on the surface $\calm_2$ in Figure \ref{2op} in the direction $(m=1, n=2)$ has a length
$4 l_p$, which is twice as long as any of the saddle-connexions 
existing in $\calm_1$. A way of understanding how this difference between the sets of 
diffractive orbits still allows isospectrality is to prove the relation (\ref{relationG})
between the Green functions, using their expression (\ref{greenstovicek}) as a definition.
This will be done in the following section, by establishing a certain
correspondence between the scattering trajectories.

\subsection{Correspondence between diffractive orbits}

Let us consider a contribution to (\ref{greenstovicek}) for any of the two surfaces. 
It is a succession of scatters and saddle-connexions of lengths 
$(k_{p_1} l_{p_1}, k_{p_2} l_{p_2}, ..., k_{p_n} l_{p_n})$, 
where the $k_i$ are integers and the $l_i$ are defined by (\ref{lp}).
In the case of forward diffraction ($\theta_k=\pi\ [2\pi]$), the integral over $s_k$ has a pole 
at $s_k=0$. But the discontinuity arising from this singularity is canceled by an opposite discontinuity 
in the term of order $(n-1)$ in  (\ref{greenstovicek}), as it should for physical reasons
of continuity. This cancellation is discussed in \cite{HanTha03}. 
This means that in the expression (\ref{greenstovicek}) one has to
interpret any term containing a forward scattering $\theta_k=\pi\ [2\pi]$ between $a$ and $b$
as the limit for $\epsilon\to 0$ of a term corresponding to a path from $a_\eps$ to $b_\eps$,
with $a_\eps\to a$ and $b_\eps\to b$. This shifted path from $a_\eps$ to $b_\eps$ 
will have two contributions: a straight path 
from $a_\eps$ to $b_\eps$ missing the singularity, plus a sum over all the scattering contributions
starting from $a_\eps$ and winding any number of times around the singularity (see Figure \ref{abscat}).
\begin{figure}[ht]
\begin{center}
\epsfig{file=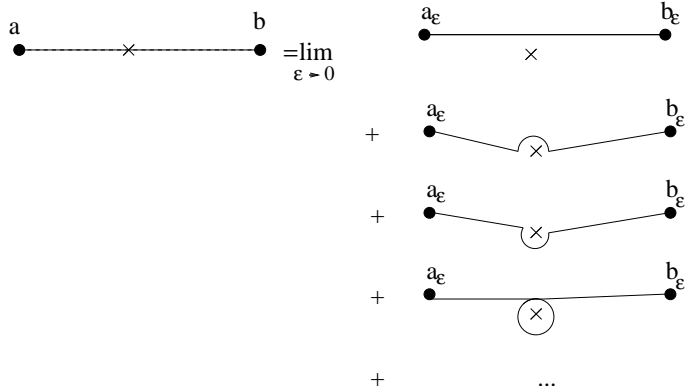,width=9cm}
\end{center}
\caption{A contribution to the Green function in case of forward diffraction}
\label{abscat}
\end{figure}

Furthermore, this still holds even if there is no scatterer at the vertex because in that case
the series of diffractive terms adds up to zero: when $\gamma_k=2\pi$,
\begin{equation}
\sum_{M_k=-\infty}^{\infty}\frac{2\pi}{(\gamma_k M_k+\pi+i s_k)^2-\pi^2}=0.
\end{equation}
Therefore any saddle-connexion of length $k l_p$ can be replaced
(as in Figure \ref{abscat} but imagining now that $a$ and $b$ are scattering vertices and that
the $\times$ corresponds to a non-diffracting vertex),
by a sum over all possible paths made of straight lines of length $k_i l_p$ 
parallel to the saddle-connexion and by 
windings around the non-scattering vertices any number of times, with the condition that $\sum k_i=k$.
So, any contribution to (\ref{greenstovicek}) can be decomposed into an infinite sum of paths 
made of a fixed number of windings around scattering vertices and any number of windings around
non-scattering vertices. This is illustrated at Figure \ref{pathinfini}.
\begin{figure}[ht]
\begin{center}
\epsfig{file=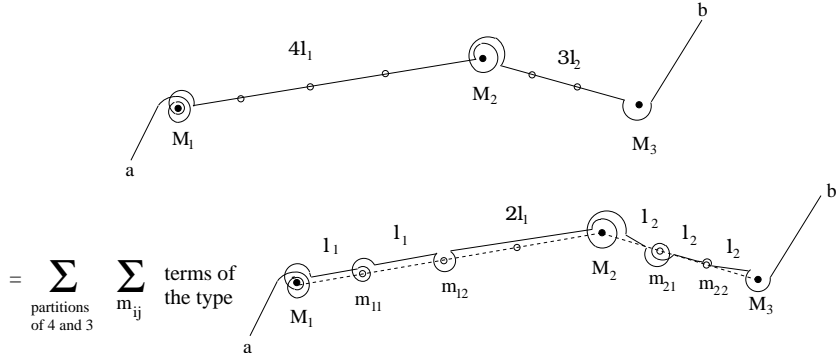,width=11cm}
\end{center}
\caption{A contribution to the Green function in case of forward diffraction. The filled circles 
are scattering vertices, the empty ones are non-scattering vertices.}
\label{pathinfini}
\end{figure}

An orbit on the torus going from $a$ to $b$ consists only of a straight path between $a$ and $b$, since
there is only one vertex and it is non-diffractive. But again, one can decompose this straight path into
a sum over paths going from the vertex to itself and winding any number of times around the vertex. These
contributions will add up to zero.
\\
\\

Let us now fix two points $a$ and $b$ on the torus and consider a path of
the type described at the left-hand side of Figure \ref{pathinfini}, 
with any given number of windings around the vertices. All the contributions to
$G^{(2)}(a, i;b,j)$, $G^{(1)}(a, i;b,j)$ or $G^{(t)}(a;b)$ are of this type. 
Reciprocally, one can write each Green function as a sum over such contributions weighted
by 1 if such a path exists between $(a,i)$ and $(b,j)$ and by 0 otherwise.

Since such a path does not hit any vertex, the sequence of matrices $(A_1,..., A_n)$ 
corresponding to the movement is well-defined (see section \ref{principeessentiel}), 
and one can construct the corresponding movement matrix $M=A_1 A_2... A_n$.
This path always exists on the torus, which means that it is always a contribution to
$G^{(t)}(a,b)$. According to section \ref{principeessentiel}, it 
will exist on the surface $\calm_\nu$, $\nu=1,2$, if and only if $a$ and $b$ are on tiles 
$i$ and $j$ such that $M^{(\nu)}_{i,j}=1$. Therefore any such path contributes to the
right-hand side of Equation (\ref{relationG}) with a weight -1/2 coming from $G^{(t)}(a,b)$
and with a weight $\sum(1/2)(T_{i, i'}T_{j,j'})$, where the sum runs over all the pairs 
$i', j'$ such that the orbit
exists between tile $i'$ and tile $j'$, i.e. such that $M^{(1)}_{i',j'}=1$. The total
weight associated to this path in the right-hand side of Equation (\ref{relationG}) is
therefore
\begin{equation}
\label{poidsdroite}
\frac{1}{2}\sum_{i', j'}T_{i, i'}T_{j,j'}M^{(1)}_{i',j'}-\frac{1}{2}.
\end{equation}
Using the commutation relation (\ref{commutationM}), we have
\begin{equation}
\label{eq26}
\sum_{i', j'}T_{i, i'}T_{j,j'}M^{(1)}_{i',j'}=\left(M^{(2)} T^2\right)_{i,j}.
\end{equation}
But it can be computed directly
from the expression (\ref{matriceT}) for $T$ that $(T^2)_{ij}=1+2\delta_{ij}$. Therefore
\begin{equation}
\label{eq27}
\left(M^{(2)} T^2\right)_{i,j}=\sum_{k}M^{(2)}_{i,k}+2 M^{(2)}_{i,j}.
\end{equation}
Since $M^{(2)}$ is a permutation matrix, the sum over a column is equal to $1$. We get, from
equations (\ref{eq26}) and  (\ref{eq27}),
\begin{equation}
\label{identite}
\sum_{i', j'}T_{i, i'}T_{j,j'}M^{(1)}_{i',j'}=1+2 M^{(2)}_{i,j}.
\end{equation}
The weight (\ref{poidsdroite}) of a path in the right-hand side of Equation (\ref{relationG}) is
therefore equal to $M^{(2)}_{i,j}$, which is exactly its weight 
in the expression of $G^{(2)}(a,i;b,j)$. This is an alternative way of proving the equality (\ref{relationG}),
using the formula (\ref{greenstovicek}) for the Green function instead of the isospectrality
of the surfaces. 
\\

In fact, Equation (\ref{identite}) provides a more precise result: it states
 that there is a ``one-to-three'' correspondence between the diffracting orbits of $\calm_2$ 
and those of $\calm_1$. The main point is that for $(i,j)$ given tiles on $\calm_2$,
$\sum_{i', j'}T_{i, i'}T_{j,j'}M^{(1)}_{i',j'}$ is the number of times a path 
of movement matrix $M^{(1)}$ appears among the 9 Green functions $G^{(1)}(a,i';b,j')$
with $i'$ pre-image of $i$ and $j'$ pre-image of $j$.
So if a path exists on $\calm_2$ between tile $i$ and tile $j$ (which means $M^{(2)}_{i,j}=1$),
then 3 copies of it exist in $\calm_1$ (and this will necessarily be between
the 3 pre-images of $i$ and the three pre-images of $j$), whereas if
a path does not exist in $\calm_2$, then by Equation (\ref{identite}) 
only 1 copy of it exists in $\calm_1$ (between
one of the three pre-images of $i$ and one of the three pre-images of $j$).

\section{Back to periodic orbits}

A final question one might want to ask is the following: why is there a one-to-one
correspondence between periodic orbits, whilst each periodic orbit in $\calm_2$
should correspond to 3 orbits in $\calm_1$? The answer is that these 3 orbits are not 
necessarily periodic.
Consider a periodic orbit in $\calm_2$, not hitting any vertex
(almost all of them verify this condition), going from tile $i$ to itself with a 
movement matrix $M^{(2)}$. Let us call $\{i_1, i_2, i_3\}$ the three pre-images of $i$.
Then, according to what we said in the previous section,
there are 3 copies of this orbit in $\calm_1$ going from 
$k\in\{i_1, i_2, i_3\}$ to $k'\in\{i_1, i_2, i_3\}$.  But these copies in $\calm_1$
are not periodic if $k\neq k'$. Moreover, we know that there are in fact 7 copies 
of this orbit in $\calm_1$, if we do not restrict ourselves to the pre-images of $i$ and $j$
as starting and ending tiles. Therefore for periodic
orbits the correspondence is more global: the condition of periodicity imposes that
we take into account not only orbits from pre-images to pre-images, but orbits from
any tile to any tile. Then, as we already said,
the number of such orbits is $\tr M^{(1)}$, which is equal to $\tr M^{(2)}$. We should 
therefore speak of a  $\tr M^{(1)}$-to-$\tr M^{(1)}$ correspondence between periodic
orbits, the value of $\tr M^{(1)}$ depending on the periodic orbit 
considered.

\section{Conclusion}
For the pair of billiards given in Figure \ref{structures}, which are made of rectangular tiles glued
together, there exists a set of neighbour matrices 
$\{R^{(\nu)}$, $L^{(\nu)}$, $\Ur^{(\nu)}$, $\Ul^{(\nu)}$, $\Dr^{(\nu)}$, $\Dl^{(\nu)}\}$
describing the way these tiles are glued together. The essential feature which accounts both for
isospectrality and for the correspondence between paths on the surface is the existence
of a ''transplantation matrix'' $T$ which has 3 properties:
\begin{itemize}
\item it is invertible (otherwise one of the spectra would just be a subset of the other)
\item it is not a permutation matrix itself (otherwise the two domains would just be
congruent)
\item it has the commutation property $T M^{(1)}=M^{(2)} T$ for all neighbour matrices $M$, 
which assures that smoothness at all the segments between the tiles and boundary conditions
will be satisfied.
\end{itemize}

It is therefore possible to generalise the previous analysis to all the pairs of isospectral
billiards made out of base tiles glued together, provided one can find a transplantation
matrix $T$ having these 3 properties. It turns out that all the examples of pairs of isospectral
billiards, as far as is known (see \cite{BusConDoy94} and \cite{OkaShu01}), are constructed by the same
application of a theorem  by Sunada \cite{Sun85} and consist of tiles of a given base shape
glued together. Moreover, Sunada's theorem implies the existence of a transplantation matrix
in each case \cite{OkaShu01}. Therefore Equation (\ref{eq26}) always holds, and 
the same arguments can be adapted to establish a correspondence 
between diffractive orbits in all these other known cases. However, the correspondence depends
on the entries of the matrix $T$ and $T^2$, and might be more complicated in the general case.
The same results could also be worked out for pairs of isospectral billiards which would not
be based on Sunada's theorem but would nevertheless have a transplantation matrix. It is not
known however if such billiards exist \cite{OkaShu01}.

\section*{Acknowledgments}
J. H. Hannay is warmly thanked for his help all along this project.

\appendix

\section{Appendix}
\label{matricesRUU}

The matrices that describe the gluings between the plates of the translation surfaces
$\calm_1$ and $\calm_2$ are obtained by Figure (\ref{structures}). For the structure $\calm_1$ 
they read
\begin{eqnarray}
\label{matrices}
&R^{(1)}=\left(
\begin{array}{ccccccc} 0&0&0&0&0&1&0\cr 0&0&1&0&0&0&0\cr 0&0&0&0&0&0&1\cr
0&0&0&1&0&0&0\cr 1&0&0&0&0&0&0\cr 0&0&0&0&1&0&0\cr 0&1&0&0&0&0&0\cr\end{array}\right), 
\Ul^{(1)}=\left(
\begin{array}{ccccccc} 0&0&1&0&0&0&0\cr 0&1&0&0&0&0&0\cr 1&0&0&0&0&0&0\cr
0&0&0&0&0&1&0\cr 0&0&0&0&1&0&0\cr 0&0&0&1&0&0&0\cr 0&0&0&0&0&0&1\cr\end{array}\right),\nonumber\\
&\Ur^{(1)}=\left(
\begin{array}{ccccccc} 1&0&0&0&0&0&0\cr 0&1&0&0&0&0&0\cr 0&0&1&0&0&0&0\cr
0&0&0&0&0&1&0\cr 0&0&0&0&0&0&1\cr 0&0&0&1&0&0&0\cr 0&0&0&0&1&0&0\cr\end{array}\right).
\end{eqnarray}

For the structure $\calm_2$ 
they read
\begin{eqnarray}
\label{matricesRU}
&R^{(2)}=\left(
\begin{array}{ccccccc} 0&0&0&0&1&0&0\cr 0&0&0&0&0&0&1\cr 0&1&0&0&0&0&0\cr
0&0&0&1&0&0&0\cr 0&0&0&0&0&1&0\cr 1&0&0&0&0&0&0\cr 0&0&1&0&0&0&0\cr\end{array}\right), 
\Ul^{(2)}=\left(
\begin{array}{ccccccc} 0&0&1&0&0&0&0\cr 0&1&0&0&0&0&0\cr 1&0&0&0&0&0&0\cr
0&0&0&0&0&1&0\cr 0&0&0&0&1&0&0\cr 0&0&0&1&0&0&0\cr 0&0&0&0&0&0&1\cr\end{array}\right), \nonumber\\
&\Ur^{(2)}=\left(
\begin{array}{ccccccc} 0&0&0&1&0&0&0\cr 0&1&0&0&0&0&0\cr 0&0&0&0&0&1&0\cr
1&0&0&0&0&0&0\cr 0&0&0&0&1&0&0\cr 0&0&1&0&0&0&0\cr 0&0&0&0&0&0&1\cr\end{array}\right).
\end{eqnarray}

One can verify explicitly that the commutation relations (\ref{commutation}) hold for these
matrices.
\\
\\
It is of some mathematical interest to notice that the groups $\Gamma^{(1)}$ and $\Gamma^{(2)}$
generated by the neighbour
matrices $\{R^{(\nu)}, L^{(\nu)}, \Ur^{(\nu)}, \Ul^{(\nu)}, \Dr^{(\nu)}, \Dl^{(\nu)}\}$ for $\nu=1,2$,
 are subgroups of $S_7$ (the group of permutations of 7 elements) of order 168. These groups 
turn out to be isomorphic to the linear group $L_2(7)$ (also called $PSL(2,7)$), which is 
the group of automorphisms of the finite projective plane of order 2, or Fano plane (see
\cite{Kar76} for a definition). The matrix $T$ is nothing but the incidence matrix of the 
graph corresponding to the Fano plane.

%%%%%%%%%%%%%%%%%%%%%%%%%%%%%%%%%%%%%%%%%%%%%%%%%%%%%%%%%%%%%%%%%%%%%%%%
%                  REFERENCES BIBLIOGRAPHIQUES
%%%%%%%%%%%%%%%%%%%%%%%%%%%%%%%%%%%%%%%%%%%%%%%%%%%%%%%%%%%%%%%%%%%%%%%%

\end{document}